\documentclass[twocolumn,amsmath,prx,amssymb,superscriptaddress,longbibliography]{revtex4-2}
\usepackage{physics}
\usepackage{graphicx}% Include figure files
\usepackage{dcolumn}% Align table columns on decimal point
\usepackage{bm}% bold math
\usepackage{qcircuit}
\usepackage{subfigure}
\usepackage{subfloat}
\usepackage{multirow}
\usepackage[figuresright]{rotating}
\usepackage{url,hyperref}
\usepackage{xcolor,amsthm,amsmath,amsxtra,amsfonts,dsfont,graphicx}
\hypersetup{
    colorlinks=true,
    linkcolor=blue,
    filecolor=blue,      
    urlcolor=blue,
    citecolor=blue
    }

\newcommand{\revise}[1]{\textcolor{black}{#1}}

\usepackage{appendix}
\usepackage{color}
\usepackage[english]{babel}

\begin{document}

%\title{Accurate and efficient quantum computations of molecules using Daubechies wavelet basis set}

\title{Accurate and Efficient Quantum Computations of Molecular Properties Using Daubechies Wavelet Molecular Orbitals: A Benchmark Study against\\ 
Experimental Data}
%{Accurate and efficient quantum computations of molecular properties using
%  Daubechies wavelet molecular orbitals: a benchmark study against experimental data}

\author{Cheng-Lin Hong}
\thanks{Contributed equally to this work.}
\affiliation{Department of Physics and Center for Theoretical Physics, National Taiwan University, Taipei 10617, Taiwan}

\author{Ting Tsai}
\thanks{Contributed equally to this work.}
\affiliation{Department of Physics and Center for Theoretical Physics, National Taiwan University, Taipei 10617, Taiwan}

\author{Jyh-Pin Chou}
\thanks{Contributed equally to this work.}
\affiliation{Department of Mechanical Engineering, City University of Hong Kong, Kowloon, Hong Kong SAR 999077, China}
\affiliation{Department of Physics, National Changhua University of Education, Changhua 50007, Taiwan}

\author{Peng-Jen Chen}
\affiliation{Department of Mechanical Engineering, City University of Hong Kong, Kowloon, Hong Kong SAR 999077, China}
\affiliation{Hong Kong Institute for Advanced Study, City University of Hong Kong, Kowloon, Hong Kong SAR 999077, China}

\author{Pei-Kai Tsai}
\affiliation{Department of Physics and Center for Theoretical Physics, National Taiwan University, Taipei 10617, Taiwan}

\author{Yu-Cheng Chen}
\affiliation{Department of Mechanical Engineering, City University of Hong Kong, Kowloon, Hong Kong SAR 999077, China}

\author{En-Jui Kuo}
\affiliation{Department of Physics and Center for Theoretical Physics, National Taiwan University, Taipei 10617, Taiwan}
\affiliation{Department of Physics and Joint Quantum Institute, University of Maryland, College Park, MD 20742 USA}

\author{David Srolovitz}
\affiliation{Hong Kong Institute for Advanced Study, City University of Hong Kong, Kowloon, Hong Kong SAR 999077, China}
\affiliation{Department of Mechanical Engineering, The University of Hong Kong, Pokfulam, Hong Kong SAR 999077, China}
\author{Alice Hu}
\affiliation{Department of Mechanical Engineering, City University of Hong Kong, Kowloon, Hong Kong SAR 999077, China}
\affiliation{Department of Materials Science and Engineering, City University of Hong Kong, Kowloon, Hong Kong SAR 999077, China}

\author{Yuan-Chung Cheng}
\affiliation{Department of Chemistry, National Taiwan University, Taipei 10617, Taiwan}
\affiliation{Center for Quantum Science and Engineering, National Taiwan University,
Taipei 10617, Taiwan}
\affiliation{Physics Division, National Center for Theoretical Sciences, Taipei, 10617, Taiwan}
\author{Hsi-Sheng Goan} \email{goan@phys.ntu.edu.tw}
\affiliation{Department of Physics and Center for Theoretical Physics, National Taiwan University, Taipei 10617, Taiwan}
\affiliation{Center for Quantum Science and Engineering, National Taiwan University,
  Taipei 10617, Taiwan}
\affiliation{Physics Division, National Center for Theoretical Sciences, Taipei, 10617, Taiwan}

\date{\today}

\begin{abstract}
  Although quantum computation is regarded as a promising numerical method for
  computational quantum chemistry, current applications of
  quantum-chemistry calculations on
 % noisy intermediate-scale quantum (NISQ)
  quantum
  computers are limited to
  small molecules. This limitation can be ascribed to  technical
  problems in building and manipulating more quantum bits (qubits) and
  the associated complicated operations of quantum gates in a quantum
  circuit when the size of the molecular system becomes large.  As a
  result, reducing the number of required qubits is necessary to make
  quantum computation  practical. Currently, the minimal STO-3G basis
  set is commonly used in benchmark studies because it requires the minimum number of  spin  orbitals. Nonetheless, the accuracy
  of using STO-3G is generally low  and thus can not provide useful
  predictions. 
  Herein, we propose to adopt Daubechies wavelet functions as an accurate and efficient
  method for quantum computations of molecular electronic 
  properties. We demonstrate that a minimal basis set constructed from
  Daubechies wavelet basis 
  can yield accurate results through a better description of the
  molecular Hamiltonian, while keeping the number of spin
  orbitals minimal.  With the improved Hamiltonian through Daubechies wavelets, we
  calculate vibrational frequencies for H$_2$ and LiH using 
  quantum-computing algorithm to show that the results are in excellent
  agreement with experimental data. As a result, we achieve quantum calculations in
  which accuracy is comparable with that of the full
  configuration interaction calculation using the cc-pVDZ basis set, whereas the
  computational cost is the same as that of a STO-3G
  calculation. Thus, our work provides a
  more efficient and accurate representation of the molecular
  Hamiltonian for efficient quantum computations of molecular systems,
  and for the first time demonstrates that predictions in agreement
  with experimental measurements are possible to be achieved with
  quantum resources available in near-term quantum computers. 

  %Quantum computation is regarded as a promising numerical method in computational quantum chemistry. Through the manipulation of the quantum bits that exploit the exotic quantum mechanical properties, a quantum computer is able to process a much greater volume of data. Thus, systems with high complexity that can hardly be handled on classical computers now become tractable on quantum computers. Because of this, an exact solution to a quantum system can be better approximated by considering more electronic configurations. On the other hand, it is of equal importance to choose an appropriate basis set in the calculations of the Hartree-Fock reference state. In this work, we demonstrate that density functional calculations with the Daubechies wavelet basis set can yield more accurate results for its better description of the reference state when compared with the commonly used STO-3G basis. In more detail, the calculations with the Daubechies wavelet basis set can be more accurate than the 6-31G full configuration interaction calculations and as accurate as the cc-pVDZ/FCI calculations, while the computational costs is roughly the same as using STO-3G basis set. Thus, our work provides an alternative option for an accurate and efficient quantum computation.

\end{abstract}

%\keywords{Suggested keywords}%Use show-keys class option if keyword
                              %display desired
\maketitle

%\section{Introduction}
\section{INTRODUCTION}

Quantum computation has recently emerged as a revolutionary numerical
scheme for computational quantum
chemistry~\cite{Kassal:2011ed,cao_quantum_2019,mcardle_quantum_2020,Bauer:2020ix},
which plays crucial roles in the design of novel materials and
drugs. Quantum simulation
has been successfully applied to the determination of the structural and electronic
properties of small molecules
~\cite{aspuru-guzik_simulated_2005,du_nmr_2010,wang_quantum_2015,kandala_hardware-efficient_2017,colless_computation_2018,grimsley_adaptive_2019,bian_quantum_2019,nam_ground-state_2020},
and is considered as the most promising near-term and real-world
applications of quantum computation. Specifically, quantum algorithms for
solutions of molecular energies have the unique ability to
provide exact results even for chemical
systems with high computational complexity. On classical computers,
the complexity of exact quantum chemical computations scales exponentially
with system size. As a result, classical approaches such as the full
configuration interaction (FCI) method are restricted to small
molecules, and approximations, such as truncated configuration
interactions or the coupled-cluster (CC) approach, are required to
reduce the computational complexity to make them tractable on
classical computers. Even so, such accurate quantum-chemistry methods are still
not practical for molecules of medium or large size. Such problems of
high classical complexity represent a sweet spot for quantum computing in
quantum chemistry to demonstrate quantum advantage.

The advantage of applying quantum-computing algorithms to
computational quantum chemistry hangs on a significant reduction in
the computational complexity; the computational complexity of 
quantum-computing algorithms for quantum chemistry grows in a polynomial
manner with increasing system size, and thus classically intractable
computations become tractable on quantum computers~\cite{Kassal:2011ed}.
%Although
At the moment, the number of quantum bits (qubits) and the length (depth) of
the quantum circuits that can be reliably performed in current noisy
intermediate-scale quantum (NISQ) devices \cite{preskill_quantum_2018}
%physical quantum computing devices
have limited the size of molecules that can be calculated. 
% are not sufficient for quantum chemical computations of large molecules,
As the power, quality and capability of quantum
computers have improved significantly 
in the last few years and will continue to advance rapidly,
one can expect that the quantum-computing
%(or hybrid quantum-classical)
algorithms will become a powerful tool for computational quantum
chemistry.
Nevertheless, innovative methods that can reduce the quantum resources
required for quantum simulations of molecular systems are crucial in
order to provide solutions to large molecules.

Among the quantum algorithms for molecular properties, the quantum phase estimation (QPE) algorithm~\cite{nielsen_quantum_2010,mcclean_exploiting_2014,babbush_encoding_2018,berry_improved_2018} and the variational quantum eigensolver (VQE) algorithm~\cite{kandala_hardware-efficient_2017,peruzzo_variational_2014,wecker_progress_2015,omalley_scalable_2016,mcclean_theory_2016}
are  currently the two most promising methods for solving electronic
structure problem in quantum-computational chemistry.
Researches in recently years have shown that implementing the QPE algorithm
typically requires a large number of quantum operations,
and hence it may require a future quantum computer with quantum error
correction to solve chemical problems for large molecules. Therefore, in this work we focus
on the VQE method, a hybrid quantum-classical algorithm, for near-term NISQ hardware. 
% are hybrid quantum-classical algorithms.
% either using a quantum phase estimator~\cite{Niel,McCl,Berr,Babb} or a variational quantum eigensolver (VQE)~\cite{Kand,Peru,Weck,OMal,McC1}.
The process of performing a quantum computation on chemical systems
% for near-term quantum hardware using the variational quantum eigensolver (VQE)
%~\cite{Kand,Peru,Weck,OMal,McC1},a hybrid quantum-classical algorithm,
using the VQE method can be described in the following steps:
(1) Perform a computation on a classical computer using traditional
quantum-chemistry methods to prepare the reference Hartree-Fock (HF)
state. (2) Construct a many-body Hamiltonian in the second
quantization form based upon single-particle solutions obtained from
the previous step. (3) Choose an \textit{ansatz}, such as the
chemically inspired ansatz of the unitary coupled-cluster with single
and double excitations (UCCSD) or the heuristic hardware-efficient ansatz with
parameterized gates, to express the many-body wave function,
resulting in a designed quantum circuit with adjustable variational  parameters.
(4) Perform the calculation and then the measurement of the parameterized quantum circuit to
solve for
the eigenvalues of the given many-body problem, i.e.,
the total energy of the system by summing the expectation value of each term in the Hamiltonian estimated through repeated state preparation
and measurement. 
(5) Perform an energy minimization to search for the lowest eigenvalue
by tuning the adjustable parameters, which is
%most frequently
carried out on a classical computer. Updated parameters are passed
into the quantum circuit to iteratively refine the wave function until
the minimization of the energy is achieved.
%Thus the current quantum computing algorithm for computational chemistry
%are hybrid quantum-classical algorithms.
% either using a quantum phase estimator~\cite{Niel,McCl,Berr,Babb} or a variational quantum eigensolver (VQE)~\cite{Kand,Peru,Weck,OMal,McC1}.

In addition to the quantum algorithm, the choice of basis set is also
of prime importance in quantum chemistry on quantum
computers. Quantum simulation of molecular systems could be realized
in either a first quantized or second quantized form. The former one
represents the molecular Hamiltonian on a multidimensional grid of
many-electron coordinates, and hence requires a large number of qubits
to carry out the quantum computation, which is still unavailable on
today's quantum computers. Therefore, realizable quantum algorithms
for quantum chemistry depend on
the second quantized form of the molecular Hamiltonian, which is
parameterized using molecular orbitals and parameters calculated from
a classical HF calculation. As a result, the choice of the
basis set determines the quality of the constructed Hamiltonian. 
Although minimal basis sets, such as the 
STO-3G basis~\cite{hehre_selfconsistent_1969}, has
proven to be extremely valuable for developing and benchmarking new
quantum-computational
algorithms~\cite{kandala_hardware-efficient_2017,google:2020bs,mcardle_quantum_2020},
the accuracy that can be obtained
using minimal basis sets tends to be poor when compared with the
experimental data due to the incompleteness of
the chosen orbitals.
In order to most efficiently utilize quantum resources, which are scarce
in available quantum computers, it is of critical importance to
minimize the number of orbitals required while preserving the quality of
the second quantized Hamiltonian.

Currently, basis sets constructed from Slater-type orbitals (STOs) are 
widely used in classical computational chemistry (largely for their 
simplicity). To improve accuracy, high-level basis sets, such as 6-31G~\cite{ditchfield_selfconsistent_1971}
and cc-pVDZ~\cite{dunning_gaussian_1989} basis sets, are regularly
employed in computational chemistry, but at the 
cost of increased computational loading. Due to the limited size and
available circuit depth in nowadays quantum computers, quantum
computations using high-level basis sets remain intractable. 
Consequently, identifying a suitable basis set is essential for
accurate {\emph {and}} efficient quantum computation. In this regard, various
methods utilizing truncated molecular orbital space (active space) to reduce required
quantum resources have been proposed. More specifically, a reduction in
the active space based on selected natural molecular orbitals has been
proposed and applied to quantum-chemistry calculations on quantum
computers with a certain degree of
success~\cite{mcardle_quantum_2020}. It is also possible to increase
the accuracy of truncated active-space calculations with classical
postprocessing~\cite{Takeshita_PRX2020}. Nevertheless, these methods
rely on additional classical post-HF calculations that could
become intractable quickly for medium-size molecules.
 Notably, in a series of recent papers, Aspuru-Guzik and coworkers
presented an efficient approach to simulate molecular systems by using
a combination of multiresolution analysis of molecular wave functions
and pair-natural orbitals on the level of second-order
perturbation
theory~\cite{kottmann_reducing_2021,schleich_improving_2021}. 
Their work demonstrates that the qubit requirements can be
significantly reduced by adopting a preoptimized basis-set-free
representation of the molecular Hamiltonian based on wavelet-based
numerical techniques. However, although a systematic procedure to
determine a high-quality Hamiltonian is
proposed in these papers, the  multiresolution analysis (MRA) approach is treated as
a black box, and the extent of benefits gained by using the
preoptimized orbitals has not been fully elucidated.

In this work, we introduce minimal basis sets consisting of Daubechies
wavelet functions~\cite{daubechies_ten_1992} to achieve
high-quality small-size basis sets for quantum simulation of molecular
systems~\cite{Wavelet_IWQCIPML_2019}. Daubechies
wavelet basis sets
have been shown to yield accurate results with excellent computational
efficiency as compared with contracted Gaussian basis sets 
%commonly used in computational chemistry 
for \revise{large molecules}~\cite{mohr_accurate_2015,ratcliff_flexibilities_2020}, and 
we believe that applications of quantum-computing algorithms to
computational chemistry could also benefit from the adoption of
Daubechies wavelet basis sets.
%In a recent work, 
Thus, the main goal of this work is to examine the use of Daubechies wavelet basis sets
to improve the accuracy and efficiency of quantum computations of
chemical systems: the optimized molecular orbitals constructed from
Daubechies wavelet functions should yield a high-quality second-quantized Hamiltonian and 
improve the accuracy of the VQE calculations.
Moreover, it is important to critically
compare quantum-computation results with experimental values in order
to evaluate the true potential of quantum-computational methods for
quantum chemistry.
Therefore, we choose to benchmark calculated vibration frequencies for
H$_2$ and LiH with experimental values.
In the following, we present a prescription to construct Daubechies wavelet basis
for high-quality second-quantized Hamiltonians for molecular systems,
and show that a minimal basis prepared in this way leads to significantly more accurate
results in terms of molecular properties compared to experimental data.

%\section{Methodology}
\section{METHODOLOGY}

% The choice of a basis set plays a critical role in determining the
% performance of a quantum-chemistry calculation. Plane waves are
% widely used to construct a basis set for solid systems because their
% periodic nature makes them efficient for computations in periodic
% unit cells. However, the use of plane waves for isolated systems is
% inefficient because they require dealing with the vacuum region on
% equal footing with the region of interest. In this sense,  atomic
% orbitals are an intuitive choice of basis set for isolated
% systems. Nonetheless, basis sets constructed from atomic orbitals
% suffer from  non-orthogonality and hence require additional
% computations of the overlap matrices. For this reason, orthogonal
% basis sets constructed from localized orbitals have been proposed
% and are now in widespread use for quantum-chemistry
% calculations. One common example is the STOs that form the minimal
% basis set STO-$n$G \cite{hehre_selfconsistent_1969}, where $n$ is
% the number of Gaussian orbitals used. The STO-$n$G basis set is
% widely used in computational quantum chemistry today, because of the
% resulting reduction in computational costs. However, t An FCI
% computation using higher level basis sets, such as 6-31G and
% cc-pVDZ, may provide improved accuracy; however, this improvement
% comes with considerably heavier computational burden. 

\subsection{The Daubechies wavelet molecular orbitals}

The choice of a basis set plays a critical role in determining the
performance of a quantum-chemistry calculation. Atomic
orbitals are an intuitive choice of basis set for isolated
molecular systems. Nonetheless, basis sets constructed from atomic orbitals
suffer from nonorthogonality and hence require additional
computations of the overlap matrices. For this reason, orthogonal
basis sets constructed from localized orbitals have been proposed
and are now in widespread use for quantum-chemistry
calculations.

Daubechies wavelets provide an alternative option for basis sets in computational quantum chemistry~\cite{genovese_daubechies_2008,mohr_accurate_2015,ratcliff_flexibilities_2020}. In wavelet theory, there are two fundamental functions, a scaling function $\phi(x)$ and a wavelet $\psi(x)$:

\begin{equation}  \label{eq:DW-1}
\begin{split}
\phi(x) & =\sqrt{2}\sum_{j=1-m}^{m} h_j\phi(2x-j), \\
 \psi(x)& = \sqrt{2}\sum_{j=1-m}^{m} g_j\phi(2x-j),
\end{split}
\end{equation}
where the coefficients $h_j$ and $g_j = (-1)^{j}h_{-j+1}$ are  elements of
the filter characterizing the order $m$ wavelet family. 
In this work, we adopt the dual-resolution Daubechies wavelet grids as
implemented in the BigDFT 
code~\cite{genovese_daubechies_2008,mohr_daubechies_2014,genovese_daubechies_2011}. There, the
three-dimensional basis functions $\phi_{i,j,k} (\textbf{r})$ are a tensor
product of the one-dimensional scaling function on the grid ($i,j,k$)
with uniform mesh $h'$ for grid points away from the nucleus,
 \begin{equation} \label{eq:DW-2}
     \phi_{i,j,k} (\textbf{r}) = \phi(x/h'-i) \phi(y/h' -j) \phi(z/h'-k),
 \end{equation}
whereas for grid points close to the nucleus, a ``fine'' region with the same uniform mesh $h$ can be described with
 \begin{equation} \label{eq:DW-3}
     \phi_{i,j,k} (\textbf{r}) = \phi(x/h-i) \phi(y/h -j) \phi(z/h-k),
 \end{equation}
augmented by a set of seven wavelets ($\psi_{i,j,k}^v,\,v=1...7$),
 \begin{equation} \label{eq:DW-4}
    \begin{aligned}
         \psi^1_{i,j,k}(\textbf{r}) = \psi(x/h-i) \phi(y/h-j) \phi(z/h-k),\\
\psi^2_{i,j,k}(\textbf{r}) = \phi(x/h-i) \psi(y/h-j) \phi(z/h-k),\\
\psi^3_{i,j,k}(\textbf{r}) = \psi(x/h-i) \psi(y/h-j) \phi(z/h-k),\\
\psi^4_{i,j,k}(\textbf{r}) = \phi(x/h-i) \phi(y/h-j) \psi(z/h-k),\\
\psi^5_{i,j,k}(\textbf{r}) = \psi(x/h-i) \phi(y/h-j) \psi(z/h-k),\\
\psi^6_{i,j,k}(\textbf{r}) = \phi(x/h-i) \psi(y/h-j) \psi(z/h-k),\\
\psi^7_{i,j,k}(\textbf{r}) = \psi(x/h-i) \psi(y/h-j) \psi(z/h-k).
    \end{aligned}
\end{equation}
The basis functions on the fine grid and the coarse grid, which is
half as dense as the fine one, can be derived from
Eq. (\ref{eq:DW-1}). 

The advantages of Daubechies wavelets are as follows. First, wavelets
are localized in both real and reciprocal space. Localization in real
space is essential for efficient calculations of isolated
systems. Localization in reciprocal space, on the other hand, is
important for preconditioning since it provides approximate
eigenfunctions of the kinetic energy operator. 
Therefore, the Daubechies wavelet basis provides an accurate
representation of molecular Hamiltonian in
spatially localized grid points. In this regard, it is closely related
to the recently proposed plane-wave dual-basis approach for low-depth
quantum simulation of
materials~\cite{babbush_low-depth_2018,kivlichan_quantum_2018}. Second, the
completeness of the Daubechies wavelet basis set eliminates the
superposition error induced by the incompleteness of basis sets like
STO-3G. Third, the use of wavelets as basis sets 
%exhibits 
\revise{provides strong} adaptivity
in programming. 
\revise{In the particular implementation used in this work, the space is}
%can easily be 
simulated using fine  and
coarse grids. The former is used to describe the region containing the
chemical bonds and the latter for the region of exponentially decaying
wave-function tails. 
\revise{As a result, the wavelets with a reasonable grid size provide a basis higher in quality than most contracted Gaussian basis sets~\cite{daubechies_ten_1992,mohr_accurate_2015}}
These characteristics make chemical accuracy
achievable at \revise{a more} affordable computational cost. 

Performing quantum simulation of molecular energy directly on a
Daubechies wavelet basis would be intractable due to the large number
of grid points. Nevertheless, it has been demonstrated that a
self-consistent field calculation with Daubechies wavelets yields
molecular orbitals suitable for quantum-chemistry
calculations. Furthermore, a high-quality minimal basis set using low-lying
molecular orbitals constructed from Daubechies wavelet yields very accurate results
in density-functional calculations
\cite{genovese_daubechies_2008,mohr_daubechies_2014,mohr_accurate_2015,ratcliff_flexibilities_2020}. Note
that in this manner we take advantage of both the accurate
representation of the Daubechies wavelet grid method and the reduced size of a minimal basis.

A molecular orbital (MO) can be constructed from Daubechies wavelet
functions by superposition of the the integer-translated
scaling functions and wavelets centered at the grid points in the
computational domain. Therefore, a MO basis function
\(\chi(\textbf{r})\) can be expressed as
 \begin{equation}
\label{mult}
   \chi(\textbf{r})=\sum _{i,j,k}S_{i,j,k}\phi_{i,j,k}(\textbf{r})+\sum _{i,j,k}\sum _{v=1}^7D_{i,j,k}^v\psi_{i,j,k}^v(\textbf{r}),
\end{equation}
where $S_{i,j,k}$ and $D_{i,j,k}$ are the expansion coefficients. In a
typical calculation, the coefficients can be optimized iteratively in a HF
self-consistent-field (SCF) calculation, producing a large set of optimized
MOs each represented by Daubechies wavelets. A minimal number of MOs
can then be selected from these HF-SCF MOs to form a minimal basis
set. This set of \emph{minimal basis Daubechies wavelet molecular
  orbitals} is constructed in a much larger Hilbert space spanned by
the grid of Daubechies wavelet functions and also adapted to the
structure of the molecule in the HF-SCF calculation. Naturally, the
Daubechies wavelet minimal basis set thus constructed would have a
higher quality compared with conventional minimal basis set such as the STO-3G
basis, although with the same number of basis MOs.
The number of spin orbitals is
one of the main factors that determines the numerical expense of
quantum computing. Therefore, accurate quantum computations with fewer
spin orbitals (and hence less qubits and lower quantum circuit depth
needed) are highly desired for practical computational chemistry.
In this work, the Daubechies wavelet molecular orbitals are obtained by performing a
HF calculation using the BigDFT
code~\cite{genovese_daubechies_2008,mohr_daubechies_2014}, an {\it ab
  initio} software package that employs Daubechies wavelet basis
sets. From the BigDFT output, we select a minimal number of spin
orbitals to form a set of minimal basis Daubechies wavelet molecular orbitals
for the following quantum simulations.
We refer the interested readers to Ref. \cite{daubechies_ten_1992}
for a more detailed discussion of Daubechies wavelets. 

\subsection{Second-quantized Hamiltonian}

Given a set of molecular orbitals, one can construct the many-body
Hamiltonian in the following second quantization form:
\begin{equation}
H = \sum_{pq}{h_{pq}a_p^\dag
  a_q}+\frac{1}{2}\sum_{pqrs}{h_{pqrs}a_p^\dag a_q^\dag a_ra_s}, \label{eq:2nd_quantized}
\end{equation}
where
\begin{align}
h_{pq} & =  \int \mathbf{dx} \, 
\chi_p^{*}(\mathbf{x}) \Big( -\frac{\nabla^2}{2} - \frac{Z}{|\mathbf{r} - \mathbf{R}|} \Big) \chi_q(\mathbf{x}), \\
h_{pqrs} & =  \int \mathbf{dx_1} \mathbf{dx_2}\, 
\frac{\chi_p^{*}(\mathbf{x_1})\chi_q^{*}(\mathbf{x_2})\chi_r(\mathbf{x_2})\chi_s(\mathbf{x_1})}{|\mathbf{r_1} - \mathbf{r_2}|}.
\end{align}
In this study, we take advantage of the minimal basis Daubechies
wavelet molecular orbitals provided by the BigDFT code to construct
second-quantized Hamiltonians. 
In BigDFT, the MO wave function is constructed by real-space wavelets,
in which the grid spacing would control the accuracy. 
We therefore
%need to
perform convergence tests on different grid parameters
to determine the best resolution. 
\textcolor{black}{Additionally, unlike calculations with the Gaussian-type basis, our calculations treat 
the core electrons of a given molecule 
%are described by 
using GTH-HGH pseudopotentials \cite{goedecker_separable_1996,hartwigsen_relativistic_1998}. 
After selecting suitable grid parameters, with the permutation
symmetries of the electron integrals, the one-electron integrals $h_{pq}$
can be directly calculated via the BigDFT code, and we need to add a Poisson solver \cite{genovese_efficient_2006,genovese_efficient_2007} to 
%treat 
evaluate the two electron integrals $h_{pqrs}$.}

Moreover, in order to gain insights into the accuracy of the minimal basis Daubechies
wavelet method, we also carry out calculations using the STO-3G
molecular orbitals. In this case, the coefficients 
$h_{pq}$ and $h_{pqrs}$ are computed from  one- and two-electron
integrals using the built-in function in PSI4 for STO-3G calculations.

%\begin{figure}[b]
%\centerline{
%\includegraphics[width=8.5cm]{H2_circuit.png}
%\caption{Quantum circuit for H$_2$ with reduction of qubits considering the $\mathbb{Z}_2$ symmetry. $Z_\theta $ is $R_z$ %gate with $\theta$ being the variational parameter.}
%\label{Fig:circuit}}
%\end{figure}

\subsection{Quantum computation}

In order to solve for the electronic energy from the second-quantized
Hamiltonian using quantum computation, parity encoding
\cite{bravyi_fermionic_2002,seeley_bravyi-kitaev_2012} is performed to
transform the second-quantized fermionic operators in
Eq. (\ref{eq:2nd_quantized}) to qubit operations in a quantum
circuit. 
The $\mathbb{Z}_2$ symmetry is exploited to further reduce the number
of qubits \cite{bravyi_tapering_2017} required to carry out the
calculation. 
Furthermore, we apply the VQE approach to calculate molecular
energies.  If we parameterize the trial wave function  
$\ket{\Psi( \va*{\theta})}$ (\textit{ansatz}), the variational principle guarantees that \begin{equation}\label{eq: var principle}
E(\va*{\theta}) \equiv
\ev{\mathcal{H}}{\Psi( \va*{\theta})}
\ge E_0.
\end{equation}
where $E_0$ is the ground-state energy of Hamiltonian $\mathcal{H}$
evaluated by a quantum circuit.  Consequently, we use the classical
LBFGS-B optimization method  \cite{byrd_limited_1995} to update
parameters and pass into the quantum circuit to iteratively refine
the wave function to reach minimal energy.

In order to account for electronic correlations, the
many-body wave function is expressed in the unitary coupled-cluster (CC) \textit{ansatz}.
We restrict the CC expansion to single and double excitations; i.e.,
the UCCSD \textit{ansatz}
\cite{peruzzo_variational_2014,romero_strategies_2018,omalley_scalable_2016,yung_transistor_2015}
or a low-depth heuristic circuit. To benchmark the performance of the
minimal basis Daubechies wavelet method, we focus on the
better-defined UCCSD \textit{ansatz} first.
Trotterization \cite{trotter_product_1959} is applied to simplify the
excitation operators in this \textit{ansatz}. Previous studies suggest that
first-order trotterization is adequate for UCCSD calculations
\cite{tranter_comparison_2018}, and different orderings of the
operators in the Trotterized form of UCC methods
\cite{grimsley_is_2020} could be employed. We reorder the excitation operator
(descending order from classical MP2 contribution) to construct the
UCCSD \textit{ansatz}. Although the UCCSD \textit{ansatz} provides 
high-quality representation of the trial wave function, the challenge of implementing UCCSD
\textit{ansatz} on quantum computer is its steep requirement of
circuit depth which scales as
$\mathcal{O} \Big( {N_\textrm{occ} \choose 2} \times {N_\textrm{virt} \choose 2}
\times N_\textrm{qubits} \Big)$ \cite{barkoutsos_quantum_2018}, where
$N_\textrm{occ}$ ($N_\textrm{virt}$) is the number of occupied (virtual) orbitals
that 
take part in the excitation, where ${N\choose m} 
= \frac{N\cdot (N-1) \cdots (N-m+1)}{m!}$.
The huge circuit depth means that the
UCCSD \textit{ansatz} is not applicable for large molecules in the
NISQ era. Hence, we also benchmark the low-depth heuristic ansatz to
replace the UCCSD ansatz on the noisy circuit to reduce the noisy
effect. Molecular orbitals obtained by using the Daubechies wavelet
and STO-3G  
%these two 
basis sets are employed to construct the many-body Hamiltonian, which 
is solved with the aid of quantum computing, as described above. 
Calculations
with STO-3G molecular orbitals and minimal basis Daubechies wavelet molecular orbitals in the UCCSD {\it
  ansatz} are denoted as STO-3G/UCCSD and DW/UCCSD, respectively.
All the quantum computations are performed using the IBM Qiskit
quantum simulators \cite{anis_qiskit_2021}.

\section{RESULTS} \label{Sec:Results}

\subsection{Performance of the minimal basis Daubechies wavelet MOs}

We first investigate the optimized molecular structures as well as
the vibrational frequencies calculated with quantum computations for
small systems including \textcolor{black}{H$_2$, LiH and BeH$_2$}. The numbers of spin orbitals included in the calculations using STO-3G (denoted as STO-3G/UCCSD) and minimal basis Daubechies wavelet molecular orbitals (denoted as
DW/UCCSD) are 4, 10 \textcolor{black}{and 12 for  H$_2$, LiH, and BeH$_2$}, respectively, 
%and are isted 
as listed in Table \ref{table-1} together with the optimized
bond lengths and vibrational frequencies. 
Note that we choose to focus
on the bond length and vibrational frequency because they are
measurable molecular properties of which accurate experimental data
are available (also listed in Table \ref{table-1}). The required number of qubits
is generally equal to that of spin orbitals. However, the number of
qubits may be reduced by two in parity encoding with $\mathbb{Z}_2$
reduction. \textcolor{black}{Then the numbers of qubits used for H$_2$ and LiH and BeH$_2$ in our
calculations are 2, 8, and 10, respectively. }

\begin{figure*}[t]
\centering
\includegraphics[width=0.85 \textwidth]{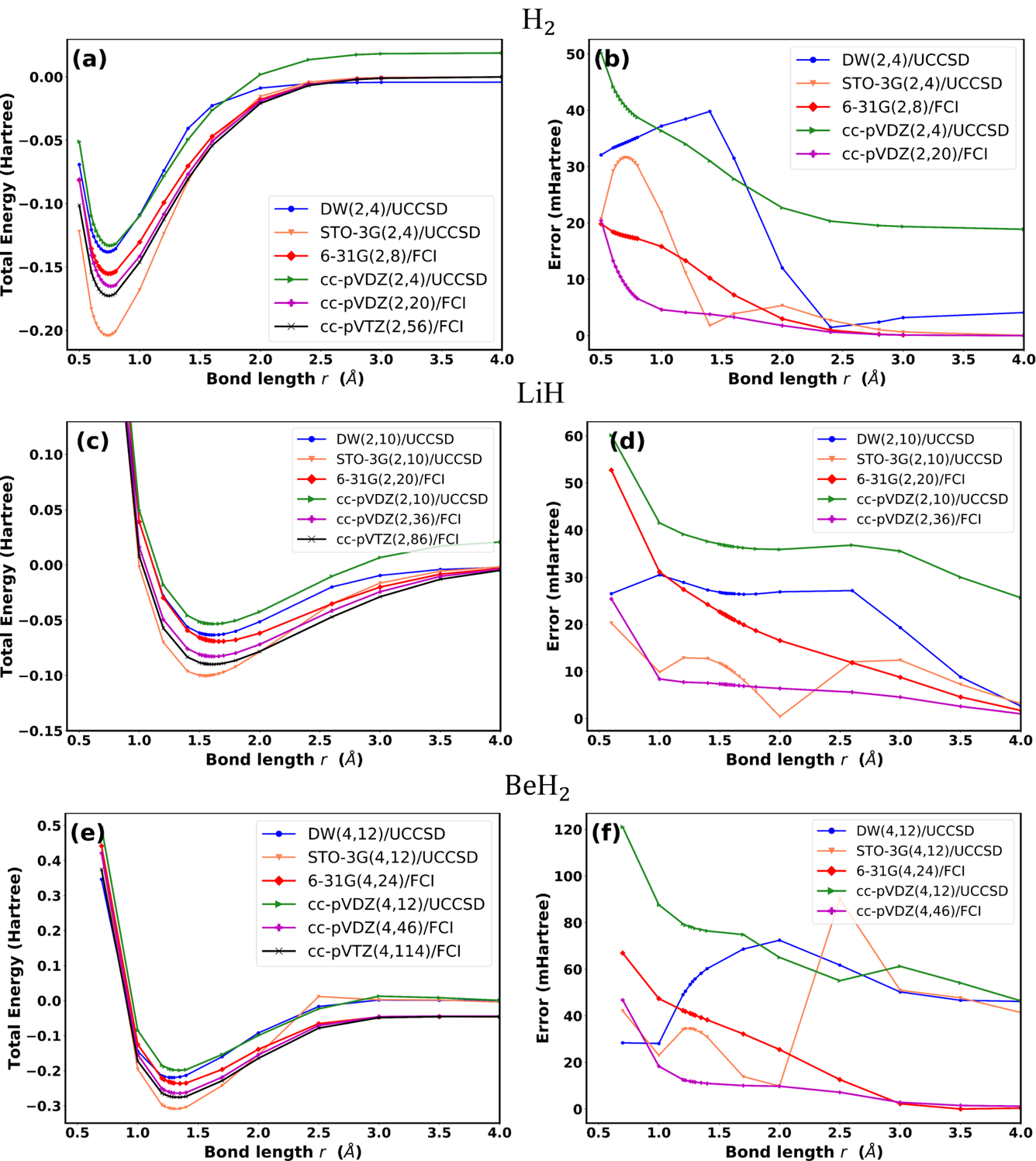}
\caption{\revise{Potential energy curves 
%(bond dissociation curves)  
in (a), (c), and (e)  and corresponding errors in (b), (d) and (f) for 
H$_2$, LiH and BeH$_2$, respectively.  The potential energies are computed with VQE on a quantum computer simulator or by exact diagonalization (FCI) on a classical computer. 
All curves are denoted with \textit{basis}$(N_e,N_\text{orb})$/method, where $N_e$ and $N_\text{orb}$ represent the numbers of electrons and spin orbitals, respectively.
The error is defined as the absolute energy difference  with respect to the ground-state energy value from the exact diagonalization (FCI) in the largest affordable Gaussian basis set of cc-PVTZ (in black curve) considered here.
%The relative error((b) H$_2$, (d) LiH, (f) BeH$_2$ ) is defined as difference (absolute value) between the exact value (FCI) for the cc-PVTZ basis set (given in black). 
The position of the minimum energy and the curvature near the minimum of the potential energy curve give the equilibrium bond length and vibrational frequency of the molecule, respectively. Note that we use the same number of spin orbitals as the  STO-3G minimal basis set for the minimal DW basis and the cc-PVTZ* basis with unoccupied orbitals frozen. }
}
\label{energy_curve}
\end{figure*}

In order to evaluate the accuracy achievable using
the minimal basis Daubechies wavelet molecular orbitals, we compare quantum simulation
results with classical FCI calculations using traditional 6-31G, \textcolor{black}{cc-pVDZ} and cc-pVTZ basis sets, denoted as 6-31G/FCI, \textcolor{black}{cc-pVDZ/FCI} and cc-pVTZ/FCI, respectively. 
For these basis
sets, FCI calculations are performed on a classical computer with the Psi4
\cite{turney_psi4_2012,parrish_span_2017,smith_p_2020} code. 
\textcolor{black}{With the frozen-core approximation, the numbers of spin orbitals used in 6-31G/FCI are 8, 20, and 24, in cc-pVDZ/FCI are 20, 36 and 46
and in cc-pVTZ/FCI are 56, 86, and 114
for H$_2$, LiH, and BeH$_2$, respectively. }
\textcolor{black}{Additionally, to clarify whether the truncated Daubechies MO wavelet basis improves accuracy over a conventional basis set with truncation in the MOs, we also consider 
a cc-pVDZ calculation with truncated unoccupied MOs (denoted as cc-pVDZ*).
%or a truncation scheme for a different basis also provides similar accuracy, we also compare cc-pVDZ with unoccupied orbitals frozen (donted as cc-pVDZ*) . 
The final numbers of orbitals used for each molecule in different basis sets are also listed in Table \ref{table-1}.}

\begin{table*}[t]
  \caption{\revise{Structural property calculations for 
 %optimization for molecules 
  H$_2$, LiH  and BeH$_2$ 
 %from the calculations 
  using different basis sets. The result from a given basis set is denoted with the abbreviation  \textit{basis/method}, e.g., the minimal Daubechies wavelet basis calculated by the method of quantum-computing UCCSD ansatz is represented by DW/UCCSD. The total number of spin orbitals used in the calculation is denoted as $n_\text{orb}$.} The equilibrium  bond lengths are obtained from the minimum energy positions and the vibrational frequencies are obtained from the curvatures near the optimized (minimum-energy) bond lengths of the potential energy curves in Figs.~\ref{energy_curve} (a), (c), and (e), respectively.  }
\begin{ruledtabular}
\begin{tabular}{lccccccccccc}
Basis set & \multicolumn{4}{c}{H$_2$} & \multicolumn{4}{c}{LiH}& \multicolumn{3}{c}{BeH$_2$}\\ \hline
& $n_\text{orb}$ & $R_\text{H-H}$ (\AA) & $\omega$ $(\text{cm}^{-1})$ & $\omega$  error  (\%)  & $n_\text{orb}$ & $R_\text{H-Li}$ (\AA) & $\omega$ $(\text{cm}^{-1})$ & $\omega$  error  (\%)  &
 $n_\text{orb}$ & $R_\text{H-Be}$ (\AA) & $\omega$ $(\text{cm}^{-1})$ \\ \cline{2-12}
STO-3G/UCCSD\footnote[1]{UCCSD: \revise{Quantum computing UCCSD ansatz method performed on  a quantum simulator}}.    & 4         & 0.74            & 4930.0                    &    12.0        & 10        & 1.55            & 1674.3           &   19.1   &  12 & 1.32  & 2273,52      \\
DW/UCCSD        & 4         & 0.74            & 4468.7                    &    1.53        & 10        & 1.61            & 1367.1           &   2.73   &  12 & 1.29  & 1948.43      \\
6-31G/FCI\footnote[2]{FCI: Classical exact diagonalization method.}       & 8         & 0.75            & 4371.6                    &    0.67        & 20        & 1.67            & 1281.7           &   8.81   &  24 & 1.35  & 1959.1      \\
\textcolor{black} { cc-pVDZ*\footnote[3]{Restricted  active space: \revise{the number of spin orbitals used is the same as that of the STO-3G minimal basis set.}}/UCCSD }  & 4  & 0.76            & 4315.2     &    1.95        & 10        & 1.62            & 1372.2           &    2.38   &12  &1.34  & $\triangle$\footnote[4]{\revise{Nonsmooth potential energy curve in the region near the optimized (minimum-energy) bond length }.}   \\
cc-pVDZ/FCI     & 20        & 0.76            & 4382.7                    &    0.42        & 36        & 1.62            & 1369.5           &   2.56       &  46 & 1.34  & 2032.1  \\
cc-pVTZ/FCI     & 56        & 0.74            & 4411.0                    &    0.22        & 86        & 1.61            & 1391.9           &   0.97       &  114 & 1.33  & 2029.7  \\
Experiment \cite{noauthor_langes_2017}       & $\cdots$        & 0.74            & 4401.2                    &      $\cdots$      	 &    $\cdots$      & 1.60            & 1405.5           &   $\cdots$       &  $\cdots$ & 1.32  & X\footnote[5]{\revise{No experimental data for the vibrational frequency of the symmetric stretch mode of  BeH$_2$.}}
\end{tabular}
\label{table-1}
\end{ruledtabular}
\end{table*}

\begin{table}[h]
\caption{\revise{Nonparallelity errors  (NPEs) in unit of mHatree for different basis sets. NPE is defined as the absolute difference between the maximum and minimal errors with respect to a reference on a given potential energy curve. The used reference here is the  largest affordable Gaussian basis set of cc-PVTZ shown in  Fig.~\ref{energy_curve}. }}
\begin{ruledtabular}
\begin{tabular}{cccc}
Basis 				& H$_2$  & LiH   &BeH$_2$\\ 
\hline
STO-3G/UCCSD     & 31.58       & 19.88    & 80.25     \\ 
DW/UCCSD         & 38.33       & 27.86    & 44.32     \\ 
6-31G/FCI        & 19.84       & 51.06    & 66.91     \\ 
cc-pVDZ*\footnote[1]{\revise{Restricted active space: the number of spin orbitals used is the same as that of the STO-3G minimal basis set.}}/UCCSD      & 31.29      & 34.47    & 74.60     \\ 
\end{tabular}
\end{ruledtabular}
\label{table-npe}
\end{table}

%The optimized structure and vibrational frequency of these molecules and the number of spin orbitals used in each calculation are listed in Table \ref{table}.

Due to the simplicity of the H$_2$ molecule, all methods give
reasonable bond lengths as compared with experimental results. On the
other hand, the vibrational frequency shows significant basis set and
computational method dependence. Clearly, the STO-3G calculations 
significantly overestimated the vibrational frequency, while the
DW/UCCSD quantum simulation yields results that are comparable in
accuracy with the high-level cc-pVTZ/FCI results. The comparison
clearly demonstrates the advantage of the Daubechies wavelet molecular
orbitals. It also shows that bond length might not be a proper
indicator for the quality of computational methods; however molecular
properties such as the vibrational frequency provides a more sensitive
indicator for the performance of the computational methods.

\revise{For LiH and BeH$_2$ , however, DW/UCCSD, cc-pVDZ*/UCCSD, cc-pVDZ/FCI and cc-pVTZ/FCI} yield bond lengths that agree well with the experimental values. 
Notably, those obtained using STO-3G are particularly poor. It is also noted that the accuracy with the assistance of quantum computing (STO-3G/UCCSD) is barely improved,
indicating the importance of choosing a proper basis set (and hence an
accurate representation of the molecular Hamiltonian). Quantum
computing cannot make up the deficiency inherited from a low-quality
basis set. Note that the number of spin orbitals used in DW/UCCSD
calculations is significantly less than that used in the cc-pVDZ/FCI  and cc-pVTZ/FCI
calculations. That means optimized molecular orbitals in the
Daubechies wavelet are very effective in achieving accurate results
with significantly reduced computational resources, making it
favorable for quantum simulations. 
%\revise{Note also that although the calculations of cc-pVDZ*/UCCSD with restricted active space provides the correct results on bond lengths, the relative errors of DW/UCCSD still can provide better relative error than cc-pVDZ/UCCSD}.

\revise{The potential energy curves (bond dissociation curves) and corresponding errors for molecules H$_2$, LiH  and BeH$_2$  are shown in
Fig.~\ref{energy_curve}.
 Representations of Hamiltonians using different basis sets for calculations are abbreviated with \textit{basis-name}$(N_e,N_\text{orb})$, where $N_e$ and $N_\text{orb}$ represent the number of electrons and spin orbitals, respectively.
The error is defined as the absolute energy difference with respect to the ground-state energy value from the exact diagonalization (FCI) in the largest
affordable Gaussian basis set of cc-PVTZ considered here.}
Since STO-3G/FCI and STO-3G/UCCSD do not exhibit significant differences, only the latter is shown and discussed.
 Inspection of Fig.~\ref{energy_curve} reveals
that the curvature around the optimized bond length seems to 
%be different, 
\revise{depend strongly on the methods used,  
implying that the presented vibrational frequency may 
%also be different. 
be a good indicator to benchmark the performance of the methods.}
Indeed, the accuracy of the calculated vibrational
frequency strongly depends on the basis set used (see Table~\ref{table-1}). STO-3G
tends to overestimate the vibrational frequency by $>10$\%. The
accuracy is better in the 6-31G/FCI calculations, albeit at higher
computational cost. Both the
H$_2$ and LiH vibrational frequencies
calculated using \revise{cc-pVDZ/FCI, cc-pVTZ/FCI} and DW/UCCSD are in excellent agreement with
the experimental values.  
\revise{ 
We can also see from Table~\ref{table-1} that the vibrational frequency of the symmetric stretch mode for BeH$_2$ calculated using DW/UCCSD is also in agreement with that obtained from  the large basis sets of cc-pVDZ/FCI  and cc-pVTZ/FCI.
Nevertheless, the later two large basis calculations require a large number of MOs, and cannot be realized on near-term quantum devices.
%sets (cc-pVDZ/FCI   and cc-pVTZ/FCI).
Note that although the calculations of cc-pVDZ*/UCCSD with a restricted active space also generate  pretty good results in  bond length and vibrational frequency for H$_2$ and LiH, their potential energy curve errors shown in  Figs.~\ref{energy_curve}(b), \ref{energy_curve}(d), and \ref{energy_curve}(f) are generally larger than those of  DW/UCCSD;
besides, the potential energy curve near the optimized equilibrium bond length of  BeH$_2$  generated by cc-pVDZ*(4,12)/UCCSD is not smooth to allow a calculation of  the  vibrational frequency, and such an issue might be originated from a convergence issue caused by the truncation in the active space. 
Similarly, in  Ref.~\cite{kottmann_reducing_2021}, Kottmann {\it et al}. showed that a
 MRA(4,12)/UpCCGSD approach that uses
% the approach of using MAR(4,12)/UpCCGSD with 
the same electron number and spin-orbital number as the minimal basis set 
%based on  multiresolution analysis (MRA) 
%in  Ref.~\cite{kottmann_reducing_2021}  
does not always converge
toward the best solution, and thus fails to produce a smooth potential energy curve for the  BeH$_2$ molecule. 
%so a more  sophisticated treatment was employed to remedy this issue. 
%Also, compare to ref. \cite{kottmann_reducing_2021}
In contrast, the minimal basis DW(4,12)/UCCSD method 
%can  permit an easier calculation of 
consistently produces a smooth potential energy curve for the BeH$_2$ molecule.
In addition, nonparallelity error (NPE), defined as the absolute difference between the maximum and minimal errors with respect to a reference on a given potential energy surface \cite{kottmann_reducing_2021,lee_generalized_2019},  could be another accuracy metric to describe the full potential energy curve properties.
One can see from Table~\ref{table-npe} that DW/UCCSD provides smaller values of NPE for the more complex molecules of LiH and BeH$_2$.}
These results, again, depict the merits of the minimal basis Daubechies wavelet molecular orbitals in calculation of
molecular properties.

\subsection{Molecular structure of water}

\begin{figure}[t]
\centering  
\includegraphics[width=8.5cm]{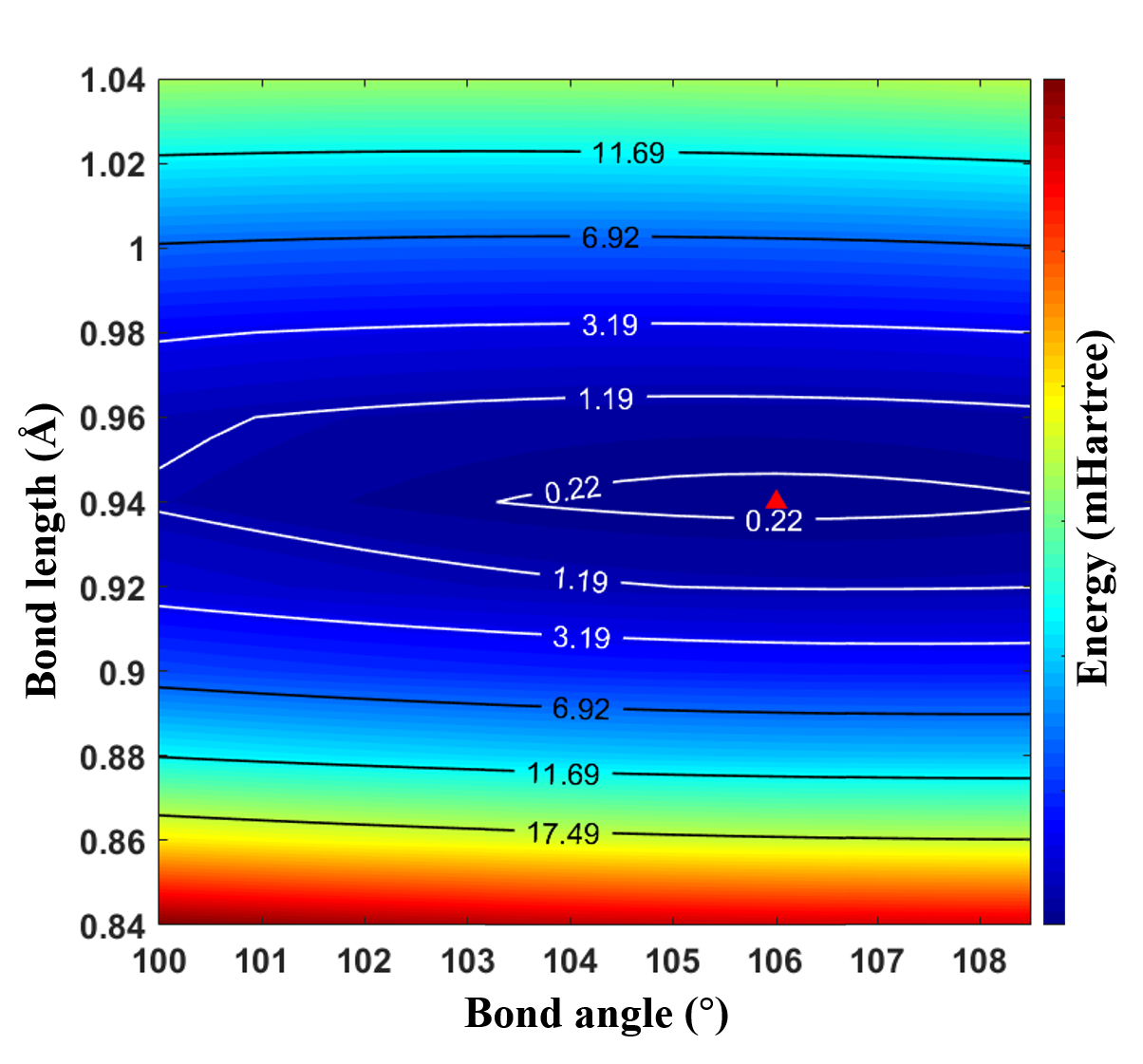}
 \caption{Two-dimensional DW/UCCSD potential energy surface of H$_2$O. Color representation of the energy is depicted by the color bar on the right and the energy of the optimized structure, marked by the red triangle, is set to zero.}
\label{Fig:water}
\end{figure}

To further investigate the accuracy of the DW/UCCSD quantum simulation
method, we apply it to compute the bond length and bond angle of the optimized
geometry of the water molecule and compare the results with those
obtained from classical methods. This is a larger molecule with a
nonlinear geometry, therefore it represents a more stringent test for
the computational methods. The computational results are listed in
Table~\ref{table:water}. In addition, for H$_2$O, a potential energy surface (PES) swept over a
range of bond lengths and  angles is displayed in
Fig. \ref{Fig:water}.
Again, the STO-3G calculations yield poor
results compared to the experimental data. In contrast, the use of minimal basis
Daubechies wavelet molecular orbitals provides significant
improvements in the predicted bond length and bond angle when compared
with the experimental values, with a prediction in bond angle even
better than the cc-pVDZ/FCI method. Note that the number of spin orbitals
used in DW/UCCSD for water is only 12,
compared to 48 used in cc-pVDZ/FCI. Clearly, the Daubechies wavelet method
significantly improves both the accuracy and efficiency for quantum
computations.  
\revise{Note that cc-pVDZ(8,12)/UCCSD in a restricted active space 
%with unoccupied orbital frozen calculates 
yields a better result in bond length and bond angle than its full CI counterpart [cc-pVDZ(8,46)/FCI], indicating the existence of a possible error cancellation that leads to the good agreement in the restricted active-space case.
In other words, the truncated cc-pVDZ(8,12)/UCCSD method yields an improved result in the H$_2$O case due to accidental error cancellation, which cannot be expected to be applicable to general molecular systems. 
In a typical atomic basis set, the calculation result is sensitive to the active-space selection, and how to select an appropriate active space for calculations is highly nontrivial.}
%still needs to be studied.}
The poor performance of the STO-3G basis is expected, since the
basis lacks the polarization functions with higher orbital angular
momentum that are required to accommodate a more flexible angular
orientations in bonding. 
On the other hand, the Daubechies wavelet
molecular orbitals \revise{constructed from large real-space grids of the wavelet functions can better capture orbital angular dependence properties, yielding a better Hartree-Fock wave function for consistent results,}
and hence provide the ability to better adapt to flexible
molecular structures. 
%For the complex molecule structure, grid-based orbitals can better capture orbitals properties, yielding better Hartree-Fock wavefunction and provide more accurate result for calculation.
 
%new part by hcl
\begin{table}[t]
  \caption{Structural property calculations of water molecule using different basis sets.  \revise{The result from a given basis set is denoted with the abbreviation  \textit{basis/method}.
%  , e.g., the minimal Daubechies wavelet basis calculated by the method of quantum computing % UCCSD ansatz is represented by DW/UCCSD. 
The total number of spin orbitals used in the calculation is denoted as $n_\text{orb}$.
The equilibrium  bond length and bond angle are obtained from the minimum-energy position coordinate of the corresponding two-dimensional potential energy surface. }}
\begin{ruledtabular}
\begin{tabular}{lccc}
Basis set    & \multicolumn{3}{c}{H$_2$O}     \\  \hline                                                                                             
             & \multicolumn{1}{c}{$n_\text{orb}$} & \multicolumn{1}{c}{$R_\text{O-H}$ (\AA)} & \multicolumn{1}{c}{H-O-H angle $(^\circ)$} \\ \cline{2-4} 
STO-3G/FCI   & 12                            & 1.03                                               & 96.7                                       \\
STO-3G/UCCSD & 12                            & 1.03                                               & 96.9                                       \\
DW/UCCSD     & 12                            & 0.94                                               & 106.0                                      \\
6-31G/FCI    & 24                            & 0.98                                               & 109.0                                      \\
\textcolor{black}{cc-pVDZ(8,12)/UCCSD}  & 12                            & 0.96                                               & 104.0                                      \\
cc-pVDZ(8,46)/FCI  & 46                            & 0.97                                               & 101.9                                      \\

Experiment \cite{noauthor_langes_2017}    &  $\cdots$                            & 0.96                                               & 104.5                                     
\end{tabular}
\end{ruledtabular}
\label{table:water}
\end{table}

Our results demonstrate that calculations of ground-state molecular
properties using minimal basis Daubechies wavelet molecular orbitals
can be as accurate as those using cc-pVDZ, but with much fewer spin
orbitals. This is especially important for quantum
simulations. For the classical segment of the calculation, it
determines the computational loading of the two-electron integral,
which is the most time-consuming part. For the quantum part, on the
other hand, the required number of qubits and the depth of the quantum
circuit depend on the number of spin orbitals, and the available
quantum resources are extremely limited, thus providing a strong
motivation to reduce the basis size. Quantum computing with the STO-3G basis set can  perform due to the small basis set size
 but its accuracy is generally low  and thus can not provide useful
  predictions.
The 6-31G basis set as well as even
larger ones require
many spin orbitals, and thus quantum computing with these basis sets at
the current stage may be impractical. We show that the
Daubechies wavelet approach
is a highly accurate and efficient alternative for the preparation of
the HF reference state that could enable quantum-computing algorithms for
larger systems. 

\subsection{Daubechies wavelet Hamiltonian and the effect of noise}

\begin{figure}[h]
\centering  
\includegraphics[width=8.5cm]{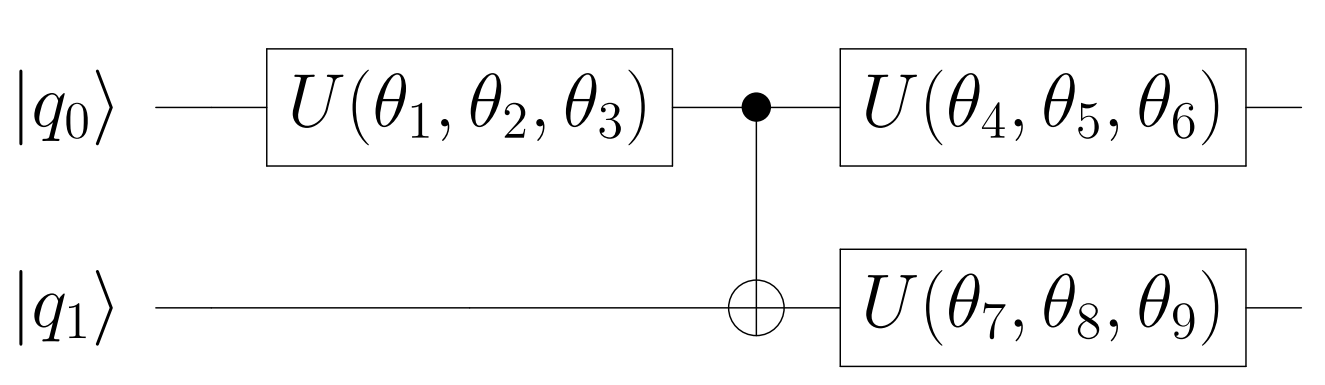}  
%\centerline{
%\Qcircuit @C=0.8em @R=1.6em {
%\lstick{\ket{q_o}} &\qw& \gate{U(\theta_1,\theta_2,\theta_3)}   & \ctrl{1} &   \gate{U(\theta_4,\theta_5,\theta_6)}  &    \qw \\
%\lstick{\ket{q_1}} &\qw& \qw  & \targ    &   \gate{U(\theta_7,\theta_8,\theta_9)}    &  		\qw
%}
%}
\caption{Two-qubit heuristic circuit  ansatz for H$_2$ in the VQE
algorithm.}
\label{fig:2Q_PQC}
\end{figure}

\begin{figure}[t]
\centering  
\includegraphics[width=8.5cm]{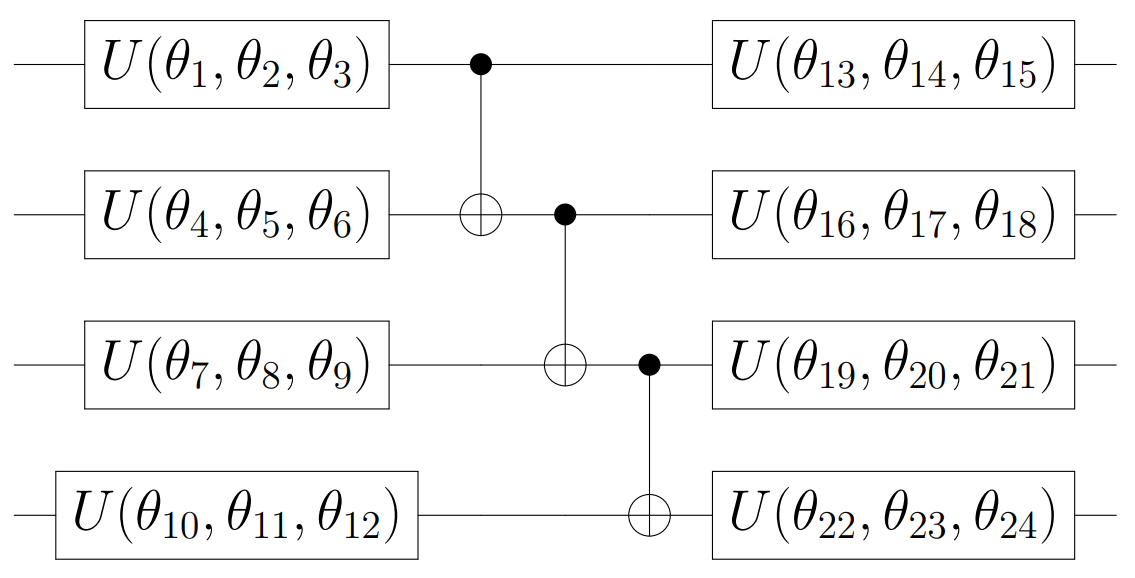}  
%\centerline{
%\Qcircuit @C=0.8em @R=1.2em {
%& \gate{U(\theta_1,\theta_2,\theta_3)}  & \ctrl{1} & \qw      & \qw        & \g%ate{U(\theta_{13},\theta_{14},\theta_{15})}  & \qw\\
%& \gate{U(\theta_4,\theta_5,\theta_6)}  & \targ    & \ctrl{1} & \qw        & \gate{U(\theta_{16},\theta_{17},\theta_{18})}  & \qw   \\
%& \gate{U(\theta_7,\theta_8,\theta_9)}  & \qw      & \targ    & \ctrl{1}   & \gate{U(\theta_{19},\theta_{20},\theta_{21})}  & \qw     \\
%& \gate{U(\theta_{10},\theta_{11},\theta_{12})}  & \qw      & \qw      & \targ   & \gate{U(\theta_{22},\theta_{23},\theta_{24})}  & \qw     \\
%}
%}
\caption{Four-qubit heuristic circuit ansatz for  LiH in the VQE
algorithm.}
\label{fig:4Q_PQC}
\end{figure}

\begin{figure*}[t]
\centering
\includegraphics[width=0.95 \textwidth]{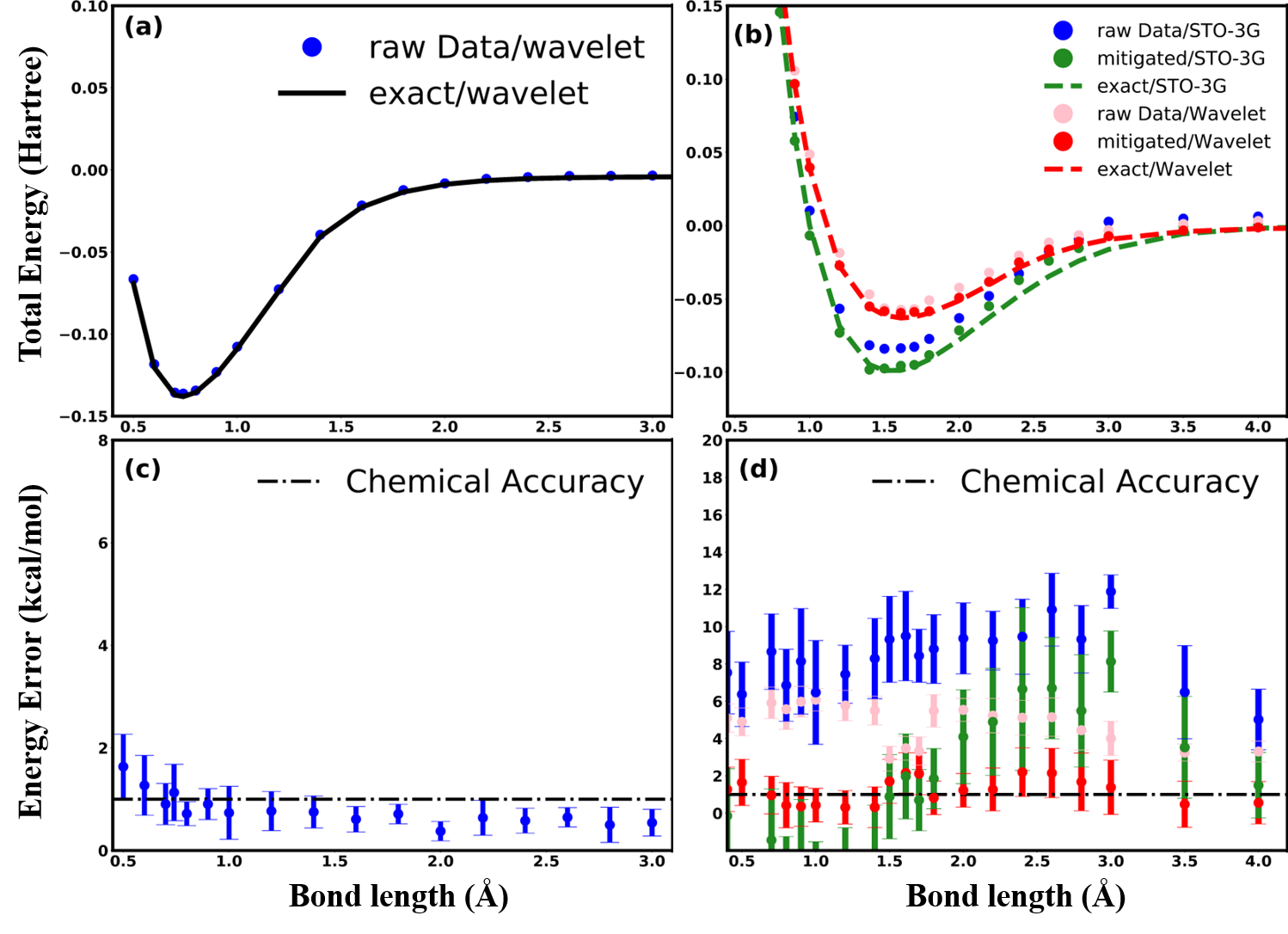}
\caption{Potential energy curves and computational errors of  H$_2$
  and LiH obtained from calculations under noisy environments. 
 \revise{The errors presented for the Daubechies
wavelet and the STO-3G methods are deviations from 
%obtained by comparing to 
their respective classical exact FCI calculation values.}
(a)  H$_2$. Results calculated using the minimal basis Daubechies wavelet
  molecular orbitals (blue dots) as well as the classical FCI (solid
  line) are shown. (b) LiH. The blue (pink) dots depict the raw data
  calculated from the STO-3G (DW) basis set. The green (red) dots
  denote the extrapolated energies over
  three different CNOT gate counts from the STO-3G (DW) basis set. These 
  simulations with noise are averaged over ten sets of 
  VQE experiments with $10^4$ shots for each iteration. The solid
  lines are the respective classical FCI results. (c) H$_2$ errors. \revise{The color scheme is the same as in the legend of (a).}
  (d) LiH errors.\revise{The color scheme is the same as in the legend of (b).} The error bars represent the 95\% confidence uncertainties that take the standard error of experiments at each CNOT gate count and the residues of linear regression into account. Most of the errors after mitigation are within the chemical accuracy defined as 1 kcal/mol.}
\label{Fig:noisy}
\end{figure*}

To investigate the degree of improvement in using a more accurate
second-quantized Hamiltonian in a quantum VQE simulation that uses a \textcolor{black}{heuristic
hardware-efficient wave-function preparation circuit \cite{kandala_hardware-efficient_2017,benedetti_parameterized_2019}}, we perform quantum
simulation using the second-quantized Hamiltonians parameterized based
on the minimal basis Daubechies wavelet molecular orbitals for H$_2$ and
LiH. We intend to compare our results with previous studies using
a hardware-efficient VQE approach based on STO-3G
Hamiltonian~\cite{kandala_hardware-efficient_2017,kandala_error_2019}, thus
we remove the 2py and 2pz orbitals of LiH to use the same number of four
qubits for the LiH system.
In addition, our numerical experiment uses low-depth
heuristic \revise{circuit} \emph{ansatzes} for hydrogen  H$_2$ (Fig. \ref{fig:2Q_PQC}) and lithium hydride LiH molecule (Fig. \ref{fig:4Q_PQC}), respectively, in the VQE
algorithm. More importantly, instead of running on a noise-free
simulator, we investigate the 
performance of the quantum simulations under
noisy quantum-computer configurations to provide a more realistic
simulation of a NISQ device. The noise model was obtained
from a realistic error statistics of the ``ibmq-santiago'' device (see
Appendix \ref{sec:noisy_quantum} for details), thus the simulations
provide a realistic benchmark of quantum computations on near-term
quantum hardware. Each energy point is the average of $10^4$ shots in each
iteration with readout-error mitigation.

Figure~\ref{Fig:noisy} shows the potential energy curves of the VQE
simulations for H$_2$ and LiH using a Hamiltonian represented by the minimal
basis Daubechies wavelet molecular orbitals. For comparison, the
results from the STO-3G basis set and respective classical FCI (labeled
exact) calculations are also presented. The H$_2$ energy curve
obtained with the Daubechies wavelet  Hamiltonian agrees well with
the exact FCI results even in the noisy simulations
(Fig.~\ref{Fig:noisy}(a)), which we attribute to the
small size of the system and the low circuit depth required for the
VQE simulation. 

The raw potential energy curve for LiH obtained with the Daubechies
wavelet Hamiltonian is presented in Fig.~\ref{Fig:noisy}(b), and it clearly deviates
from the exact FCI result, indicating the stronger influence of noise in
this larger calculation. Therefore, for LiH we also
calculate the results in higher noise 
levels and correct the obtained energy by considering the linear
zero-noise extrapolation
method~\cite{giurgica-tiron_digital_2020,larose_mitiq_2021,shee_qubit-efficient_2021} for error mitigation. The
mitigated energy curve then is in excellent agreement with the exact FCI
result. Note that the UCCSD \emph{ansatz}  for minimal basis LiH calculation
requires about over 1000 quantum gates for state preparation,
whereas the heuristic \emph{ansatz}  requires only about
11 quantum gates (Fig. \ref{fig:4Q_PQC}). Our results strongly support the effectiveness of the
low-depth heuristic \emph{ansatz} when compared to the UCCSD \emph{ansatz} (Fig.~\ref{energy_curve}).

It is also worth noting that the minimal basis STO-3G and Daubechies
wavelet molecular orbital potential energy curves are significantly
different.  While the state preparation using the low-depth heuristic
\emph{ansatz}, the Daubechies wavelet method yields results that are
in excellent agreement with experimental data (Table~\ref{table-1}); in contrast,
the STO-3G Hamiltonian yields poor results, and this cannot be
improved by additional quantum resources. Our results emphasize the
importance of using a high-quality Hamiltonian in heuristic VQE
quantum simulations.

To quantify the computational errors, the deviations from \revise{their respective classical  exact FCI 
curves} for H$_2$ and LiH are plotted in
Figs. \ref{Fig:noisy}(c) and \ref{Fig:noisy}(d), respectively. In
Fig. \ref{Fig:noisy}(c), we can see that most of the raw data achieve
chemical accuracy for the hydrogen molecule by using the Daubechies
wavelet basis set. Additionally, for LiH molecule
(Fig. \ref{Fig:noisy}(d)), calculations without error mitigation using Daubechies
wavelet molecular orbitals exhibit approximately 4 kcal/mol
error, smaller than that of the STO-3G Hamiltonian. Compared to previously published
calculations~\cite{kandala_hardware-efficient_2017,kandala_error_2019},
our simulation results demonstrate that the low-depth ansatz can be
applied to
obtain more accurate raw data by using the Daubechies wavelet basis. Furthermore,
after applying the zero-noise extrapolation, most of the mitigated
data achieve the chemical accuracy level. Note that the errors
presented for the Daubechies
wavelet and the STO-3G methods are obtained by comparing to their
respective classical FCI calculations. If comparison with experimental
data is concerned, the STO-3G data will exhibit a significant error,
as discussed earlier.
%in the previous subsection.
By using the minimal basis Daubechies
wavelet molecular orbitals to present the Hamiltonian, it is demonstrated that
more accurate results using the same quantum resources as STO-3G is
possible. Therefore, the minimal basis Daubechies wavelet molecular
orbitals provide significantly improved Hamiltonian representation,
and thus is especially adapted to
quantum simulation of molecular systems in this NISQ quantum era.

%\section{Conclusion}
\section{CONCLUSION}
\label{sec:conclusion}

In this work, we demonstrate that self-consistent field calculations
based on a Daubechies wavelet basis set can prepare a better set of
molecular orbitals, which in turn results in a better many-body
Hamiltonian. The completeness and orthogonality, in both real and
momentum space, allow Daubechies wavelet basis sets to produce
accurate and efficient representations of molecular electronic
systems. We show that quantum simulation using the
DW/UCCSD VQE method predicts accurate structures
of small molecules (H$_2$, LiH, and H$_2$O) at a
computational cost equivalent to the calculations using a minimal
basis STO-3G basis set. The improved Daubechies wavelet molecular
orbitals enable us to determine H$_2$ and LiH vibrational frequencies
with the values in excellent agreement with experimental
data. Furthermore, we show that using a Hamiltonian represented in minimal basis Daubechies
wavelet molecular orbitals together with a low-depth heuristic VQE
state-preparation circuit yield excellent results with much fewer
quantum gates compared to UCCSD ansatz. Our work indicates the
importance of the choice of basis set in applying quantum computing to
computational chemistry, and the Daubechies
wavelet basis set is shown to be highly advantageous in terms of both accuracy and efficiency.

Introducing the Daubechies wavelet basis improves the quality of the
Hamiltonian while keeping the 
quantum resources needed to implement it minimal. Furthermore, as the Daubechies wavelet molecular
orbitals are constructed from a grid of wavelet functions, it bridges the grid
method and the atomic orbital basis method for quantum computation of
quantum chemistry. As a result, the Daubechies wavelet method proposed
here yields results that are comparable to
classical high-level basis set methods with predictive power for experimental
observables. Note that the Daubechies wavelet basis offers means to systematically improve 
the accuracy of the calculations both in terms of the fineness of the
Daubechies wavelet grids used to construct the molecular orbitals and the number of
virtual orbitals used in the post-HF calculations. Thus, conventional
approaches to reduce the size of the active space as well as hybrid
postprocessing approaches to mitigate errors in quantum computers are
readily applicable to the Daubechies wavelet basis method~\cite{Takeshita_PRX2020,mcardle_quantum_2020}.

\section{Acknowledgments}
C.L.H is supported by the Young Scholar Fellowship (Einstein Program) of the Ministry of Science and Technology, Taiwan (R.O.C.) under Grants No.~MOST 109-2636-E-002-001, No.~MOST 110-2636-E-002-009, No.~MOST 111-2119-M-007-006, and No.~MOST 111-2119-M-001-004, by the Yushan Young Scholar Program of the Ministry of Education, Taiwan (R.O.C.) under Grants No.~NTU-109V0904 and No.~NTU-110V0904, and by the research project “Pioneering Research in Forefront Quantum Computing, Learning and Engineering” of National Taiwan University under Grant No.~NTU-CC-111L894605.
%C.L.H is supported by the Young Scholar
%Fellowship (Einstein Program) of the Ministry of Science and Technology in Taiwan under %Grants No. MOST 109-2636-E-002-001 and 110-2636-E-002-009, and is supported by the Yushan %Young Scholar Program of the Ministry of Education in Taiwan under Grants No. NTU-109V0904 %and NTU-110V090. 
A.H. acknowledges funding support from City University of Hong Kong under Project No.~7005615, Hong Kong Institute for Advanced Study, City University of Hong Kong, under Project No.~9360157, and Research Grants Council of the Hong Kong Special Administrative Region, China under Project No. CityU 11200120.
%C.-L. Hong, T. Cai, and J.-P. Chou contributed equally to this work.
Y.C.C thanks the Ministry of Science and Technology, Taiwan (Grants
No.~MOST 109-2113-M-002-004 and No.~MOST 109-2113-M-001-040), and National Taiwan University (Grant No. 110L890105) for financial support.
H.S.G. acknowledges support from the the Ministry of Science and Technology, Taiwan under Grants No.~MOST 109-2112-M-002-023-MY3, 
No.~MOST 109-2627-M-002-003, No.~MOST 107-2627-E-002-001-MY3, No.~MOST 111-2119-M-002-006-MY3,
No.~MOST 109-2627-M-002-002, No.~MOST 110-2627-M-002-003,
No.~MOST 110-2627-M-002-002, No.~MOST 111-2119-M-002-007, No.~MOST 109-2622-8-002-003, and 
No.~MOST 110-2622-8-002-014,
from the US Air Force Office of Scientific Research under
Award Number FA2386-20-1-4033,
and from the
National Taiwan University under Grants
No.~NTU-CC-110L890102 and No.~NTU-CC-111L894604.
J.P.C., Y.C.C. and H.S.G. acknowledge support from the Physics Division, National Center for Theoretical Sciences, Taiwan.
%(Grant No.~MOST 110-2124-M-002-012).

\appendix
%\section{Noisy quantum simulation}
\section{NOISY QUANTUM SIMULATION}\label{sec:noisy_quantum}

%In Result,
We perform the numerical simulation using the
IBM Qiskit simulator with the noise data downloaded from
the IBM Quantum 5-qubit machine ibmq-santiago
% at the date 2021-05-16. 
on 16 May 2021.
 The qubit relaxation time $T_1$, dephasing time $T_2$, and qubit frequency data are listed in Table \ref{Table 2}, the single-qubit gate error, gate length and readout error data are in Table \ref{Table 3}, and the two-qubit coupling and corresponding CNOT gate error and gate length data are listed in Table \ref{Table 4}.
Note that the IBM Quantum system service is still under active development, meaning that the noise data will change gradually

\begin{table}[hbt!]
\centering
\caption{Qubit relaxation time T1, dephasing time
T2 and qubit frequency data of the 5-qubit machine
ibmq-santiago downloaded from the IBM Quantum system service at the time when the numerical simulation is performed}

\begin{ruledtabular}
\begin{tabular}{cccc}
Qubit & T1 (µs)    & T2 (µs)    & Freq (GHz) \\ 
\hline
0     & 168.167  & 163.925 & 4.8334   \\ 
1     & 145.087  &  96.712 & 4.6236   \\ 
2     & 131.625  & 109.047 & 4.8205   \\ 
3     & 175.273  &  85.200 & 4.7423   \\ 
4     &  87.847  & 142.421 & 4.8163   \\ 
\end{tabular}
\end{ruledtabular}

\label{Table 2}
\end{table}

\begin{table}[hbt!]
\caption{
Single-qubit gate errors, and readout error data of the 5-qubit machine ibmq-santiago
downloaded from the IBM Quantum system service at the time when the experiment is performed.}
\begin{ruledtabular}
\begin{tabular}{ccc}
\label{Table 3}
Qubit & Gate error  & Readout error \\ 
\hline
0     & 0.0264\%       & 2.06\%        \\ 
1     & 0.0253\%       & 1.29\%        \\ 
2     & 0.0194\%       & 0.98\%        \\ 
3     & 0.0182\%       & 0.62\%        \\ 
4     & 0.0232\%       & 1.77\%        \\ 
\end{tabular}
\end{ruledtabular}
\end{table}

\begin{table}[hbt!]
\centering
\caption{Two-qubit CNOT gate error and gate length data of the 5-qubit machine ibmq-santiago downloaded from the IBM Quantum system service at the time when the numerical simulation is performed. The two-qubit CNOT gate error and lengths are usually different for different two-qubit coupling pairs due to different coupling strength and different qubit frequencies in the qubit pairs}
\begin{ruledtabular}
\begin{tabular}{ccc}
Coupling pair & Gate error & Gate length(ns) \\ 
\hline
{[}0, 1{]}    & 0.812\%        & 526.22         \\ 
{[}1, 0{]}    & 0.812\%        & 561.78          \\ 
{[}1, 2{]}    & 0.680\%        & 355.55          \\ 
{[}2, 1{]}    & 0.680\%         & 320.00         \\ 
{[}2, 3{]}    & 0.542\%         & 376.89         \\ 
{[}3, 2{]}    & 0.542\%        & 412.44          \\ 
{[}3, 4{]}    & 0.599\%          & 376.89        \\
{[}4, 3{]}    & 0.599\%          & 341.33         \\
\end{tabular}
\end{ruledtabular}

\label{Table 4}
\end{table}
\bibliography{reference_newQ.bib}

\end{document}